\documentclass[twocolumn,showpacs,preprintnumbers,amsmath,amssymb]{revtex4}

\usepackage{epsf}

\begin{document}

\title{Optimal time delay embedding for nonlinear time series modeling}
\author{Michael Small\thanks{Tel: +852
2766 4744 , Fax: +852 2362 8439, email: {\tt ensmall@polyu.edu.hk}.}} 
\affiliation{Department of Electronic and Information
Engineering\\ Hong Kong Polytechnic University, Hung Hom, Kowloon,
Hong Kong} 

\date{\today} 

\begin{abstract}
When building linear or nonlinear models one is faced with the problem
of selecting the best set of variable with which to predict the future
dynamics. In nonlinear time series analysis the problem is to select the
correct time delays in the time delay embedding. We propose a new
technique which can quantify the suitability of a particular set of
variables and we suggests a computationally efficient scheme to
determine the best non-uniform time delay embedding for modeling of time
series. Our results are based on the assumption that, in general, the
variables which give the best local constant model will also give the
best nonlinear model. In a wide variety of experimental and simulated
systems we find that this method produces dynamics that are more
realistic and predictions that are more accurate than standard uniform
embeddings.
\end{abstract}


\maketitle

An autoregressive model predicts future evolution as a linear
combination of past observations. An artificial neural network combines
various ``inputs'' to predict unknown data. While the theory of
modeling (both linear and nonlinear) is well developed there is no
general method to choose the correct variables (``inputs'' or
``observations'') with which to predict future dynamics. In this
communication we propose a quantitative criterion which may be used to
assess the relative merit of various combinations of variables. To
illustrate this concept we consider the specific problem of time series
prediction when only a scalar time series is available and one
reconstructs the underlying dynamics with a time delay embedding. The
generalization of this method to other situations, such as multivariate
time series, is obvious.

Very often, a high dimensional physical system is only observable through
a single scalar variable. The method of time delay embedding \cite{fT81}
is widely applied to estimate the evolution of the underlying
vector field. From a scalar time series $\{x_t\}_{t=1}^N$ of $N$
observations one reconstructs a
vector time series with evolution topologically equivalent to the
original system via the transformation
\begin{eqnarray}\label{uni}
x_t & \rightarrow & 
(x_{t},x_{t-\tau},x_{t-2\tau},\ldots x_{t-(d_e-1)\tau}).
\end{eqnarray}
Even in the restricted field of time delay embedding, there is no
general method to select the best group of variables
(\ref{uni}). Several authors have argued that the critical parameter is
the product $d_e\tau$ \cite{cC03}, but, in general one must estimate
both the embedding dimension $d_e$ and the embedding lag
$\tau$. Embedding dimension is usually estimated via the application of
geometric methods such as false nearest neighbors \cite{mK92} and
embedding lag is related to underlying time-scales (such as
pseudo-periodicity) in the time series \cite{hA96b}. However, most of
these approaches are motivated by the objective of accurately estimating
dynamic invariants. In \cite{cC03} we see that which embedding criterion
is judged to be best will depend on the adjudicating criterion.

\begin{table*}
\begin{center}
\begin{tabular}{r|c|c|c|c|c|c|c|l}
data & $N$ & $f$ & $T$ & $d_e$ & $\tau$ & $d_w$ & $d$ & $\ell_1,\ldots,\ell_k$\\\hline
Sunspots & 301 & 11.0 & 1/year& 6 & 3 & 7 & 10 & 1,2,5\\
Ventricular Fibrillation & 6000 & 27.0 & 167 Hz & 6 & 7 & 2 & 10 & 1,5,6\\
Laser & 4000 & 7.48 & 25 Mhz & 8 & 2 & 32 & 32 & $1,2,6,7,11,14,15,21,24,25,30$\\
R\"ossler & 1000 & 12.4 & 5 Hz & 3 & 3 & 6 & 10 & $1,5,7$\\
R\"ossler+noise & 1000 & 12.4 & 5 Hz & 5 & 3 & 9 & 10 & $1,2,4,5,6,7,9$\\
Lorenz & 1000 & 18.7 & 20 Hz & 5 & 43 &4 & 10 & $1,3$\\
Lorenz+noise & 1000 & 18.7 & 20 Hz& 5 & 42 & 8 & 10 & $1,2,3,5,6,7,9$\\
\end{tabular}
\end{center}
\caption{Embedding parameters for the various data considered in this
  paper: data length ($N$), physical sampling rate ($f$), data
``pseudo-period'' estimated as mean cycle time ($T$), embedding dimension computed with the method
  of false nearest neighbors ($d_e$), embedding lag estimated by the
  first zero of autocorrelation ($\tau$), non-uniform embedding window
  computed with the method described in \cite{window} ($d_w$), and the
  non-uniform set of embedding lags (such that
  $(x_{t-\ell_1},\ldots,x_{t-\ell_k})$ is used to predict $x_t$). With
  the exception of the Lorenz system, $\tau$ is approximately
  one-quarter of the pseudo-period of the time series.}
\label{table}
\end{table*}

Furthermore, there is no reason to suppose that a
{\em uniform} embedding, such as (\ref{uni}), is the correct approach
\cite{kJ98}. In general, the problem of estimating the optimum embedding
should be restated as: find the parameters $\{\ell_i | i=1,\ldots k\}$
and the embedding window $d_w$, 
where $1\leq\ell_1\leq\ell_i<\ell_{i+1}\leq\ell_{k}\leq d_w$ and the time
delay embedding 
\begin{eqnarray}
\label{nonuni}
x_t & \rightarrow & 
(x_{t-\ell_1},x_{t-\ell_2},x_{t-\ell_3},\ldots x_{t-\ell_{k}})
\end{eqnarray}
is somehow the ``best''. For nonlinear time series modeling, embedding
strategies such as this were introduced by Judd and Mees \cite{kJ98} and
are described as {\em non-uniform} embeddings.  This problem is now a
special case of the more general problem of selecting the best set of
variables (``inputs'') to model (for example, via an artificial neural
network) some unknown quantity.

Unfortunately, application of equation (\ref{nonuni}) makes the problem of
selecting embedding parameters considerably more complicated. In this
paper we propose a suitable criterion for quantitatively comparing
embedding strategies and describing an efficient scheme for the computation
of $\{\ell_i | i=1,\ldots,k\}$ and $k$. We apply this method to
several experimental and simulated time series and show that the
non-uniform embedding strategy has many advantages over the standard
techniques (\ref{uni}). Non-Uniform embedding strategies usually utilise a smaller
embedding window and provide better nonlinear predictions (small mean
prediction error). We find that employing a non-uniform embedding strategy
allows one to simulate complex nonlinear dynamics that are qualitatively
more like the true system (or in the case of experimental data, simply
more plausible).

Often, the purpose of time delay embedding is to estimate correlation
dimension \cite{effgka} or other dynamic invariants \cite{hA96b}. In
such situations, embeddings such as (\ref{uni}) are usually adequate. In
this work we focus on the problem of estimating the underlying evolution
operator of the dynamical system from a single scalar observable. We are
therefore interested in obtaining the most accurate prediction of the
observed data values. By doing so we hope to capture the long term
dynamics of the underlying system. To achieve this we adopt the
information theoretic measure {\em description length} \cite{jR89} and
seek to choose the embedding which provides the minimum description
length. This method can be applied equally to a variety of other
modeling regimes.

Roughly speaking the description length of a time series is the
compression of the finite precision data afforded by the model of that
data \cite{kJ95a}. If a model is poor then it will be more economical to
simply describe the model prediction errors. Conversely, if a model fits
the data well, then the description of that model and the (presumably
small) model prediction errors will be more compact. However, if a model
over-fits the data \cite{mdlnn} then the description of the model itself
will be too large. In \cite{window} we showed that the description
length $DL(\cdot)$ of a time series $\{x_t\}$ is approximated by 
\begin{eqnarray}
\nonumber
DL(\{x_t\}) &  \approx &
\frac{N}{2}\left(1+\ln{2\pi}\right)+
\frac{d}{2}\ln{\left[\frac{1}{d}\sum_{i=1}^d(x_i-\overline{x})^2\right]}\\
\nonumber & & \mbox{  }
+\frac{N-d}{2}\ln{\left[\frac{1}{N-d}\sum_{i=d+1}^Ne_i^2\right]}\\
 & & \mbox{  }
+d+ DL(d)+DL(\overline{x})+DL(\mathcal{P}).
\label{dl1}
\end{eqnarray}
where $d=\max_{i}\{\ell_i\}=\ell_{d_e}$, $\overline{x}=E(x_t)$ is the mean of the
data, $\{e_t\}_t={d+1}^N$ are the model prediction errors, and $DL(\mathcal{P})$ is
the description length of the model parameters. The description length
of an integer $d$ can be shown to be
$DL(d)=\lceil\log{d}\rceil+\lceil\log{\lceil\log{d}\rceil}\rceil+\ldots$
where each term on the right is an integer and the last term in the
series is $0$ \cite{jR89}. Furthermore,
$\frac{N}{2}\left(1+\ln{2\pi}\right)+DL(\overline{x})$ is independent of
the embedding strategy. Hence, the optimal embedding strategy is that
which minimizes 
\begin{eqnarray}
\nonumber
\frac{d}{2}\ln{\left[\frac{1}{d}\sum_{i=1}^d(x_i-\overline{x})^2\right]}
+d+ DL(d)+\\
\frac{N-d}{2}\ln{\left[\frac{1}{N-d}\sum_{i=d+1}^Ne_i^2\right]}
+DL(\mathcal{P}).
\label{dl2}
\end{eqnarray}
The first three terms in (\ref{dl2}) may be computed directly. However,
the last two terms require one to estimate the optimal model. 

As in \cite{window}, for the purposes of computational expediency, we
restrict ourselves to the class of local constant models. In the current
context this is not unreasonable as we hope to obtain an
embedding which spreads the data in phase space based on the
deterministic dynamic evolution. Under this assumption, 
$DL(\mathcal{P})=0$ and the model prediction error 
$\frac{1}{N-d}\sum_{i=d+1}^Ne_i^2$
may be computed via ``drop-one-out'' interpolation. That is,
$e_{i+1}=x_{i+1}-x_{j+1}$ where $j\in\{1,2,\ldots,N\}\backslash\{i\}$ is
such that $\|x_i-x_j\|$ is minimal.  Note that, in the limit as
$N\rightarrow\infty$ (i.e. $N\gg d$) optimizing (\ref{dl2}) is
equivalent to finding the embedding which provides the best prediction
(the last two terms of (\ref{dl2}) dominate).

To minimize equation (\ref{dl2}) we assume that the maximum number of
inputs, $d$, has already been calculated. One may choose $d=d_w$, the
embedding window computed using the method described in
\cite{window}. Alternatively, one may either assign an arbitrary value
for $d$ or use $d=d_e\tau$ where both $d_e$ and $\tau$ are estimated by
one of the many standard techniques. The technique suggested in
\cite{window} gives an upper bound on the embedding window $d_w$. But
the method offered in \cite{window} provides no way of choosing the
optimal set of embedding lags. In fact, the main conclusion of
\cite{window} is that estimating $d_w$ should be done prior to
modeling, but estimating the actual embedding lags should be considered
part of the modeling process. In this paper we apply the computational
procedure described below to select the embedding (\ref{nonuni}) that
optimizes (\ref{dl2}). This solves the main problem raised by
\cite{window}. For the numerical simulations presented here, we choose
$d$ such that $d\geq d_w$.

An exhaustive search on the $2^d$ possible embedding strategies is only
feasible for small $d$. For large $d$ (i.e. $d>10$) we utilise a genetic
algorithm to determine the optimum embedding strategies. Furthermore, to
reduce the computational effort in estimating the model prediction error
for large $N$ ($N>1000$) we minimize the prediction error only on a
randomly selected subset of the data. Our calculations show that neither
of these approximations adversely affect our results. The results of the
genetic algorithm are robust and accurate. Furthermore, we find that
provided the data subset is selected {\em with} replacement and that it
is moderately large, the final solution is independent of the specific
subset selected.  Our choice of genetic algorithm (over alternative
optimisation techniques) is arbitrary. Other techniques (such as
simulated annealing) may also perform well, but remain untested.

We tested this algorithm with data from three experimental systems (the
famous annual sunspot time series, a chaotic laser \cite{eW93}, and a
recording of human electrocardiogram during ventricular fibrillation (VF)
\cite{cic4}), and two computational simulations (R\"ossler and
Lorenz equations) both with and without the addition of Gaussian noise
with a standard deviation of $5\%$ that of the data. For each data set
we estimated the embedding window $d_w$ \cite{window}, the embedding
dimension $d_e$ (via false nearest neighbors) and the embedding lag
$\tau$ (using the first zero of the autocorrelation). The results of these
calculation together with the non-uniform embedding strategy estimated using
the methods proposed here are reported in Table \ref{table}.

\begin{table}
\begin{center}
\begin{tabular}{r|c|c|c|c|c}
 &\multicolumn{5}{c} {correlation dimension}\\\cline{2-6}
 &\multicolumn{2}{c|} {uniform} &\multicolumn{2}{c|} {non-uniform} & data \\\cline{2-5}
data & median & mean & median & mean & est.\\\hline
Sunspots & 2.09 & 3.04$\pm$4.95 & 2.24 & 2.19$\pm$0.413 & 1.89\\
VF & 1.46 & 1.46$\pm$0.0486 & 1.56 & 1.56$\pm$0.0387 & 1.62\\
Laser & 2.21 & 2.20$\pm$0.116 & 2.48 & 2.45$\pm$0.190 & 2.13\\
R\"ossler & 1.31 & 1.32$\pm$0.128 & 1.34 & 1.34$\pm$0.0969 & 1.59\\
R.+noise & 1.96 & 1.94$\pm$0.135 & 1.89 & 1.86$\pm$0.141 & 1.82\\
Lorenz & 1.98 & 1.98$\pm$0.0789 & 1.86 & 1.86$\pm$0.0968 & 1.97\\
L.+noise & 1.64 & 1.64$\pm$0.0743 & 1.64 & 1.63$\pm$0.0576 & 1.61\\
\end{tabular}
\end{center}
\caption{Comparison of correlation dimension estimates for the data and
local constant model simulations (as described in \cite{cyclsurr}) using
either the standard uniform or non-uniform embedding strategy. We computed
$30$ simulations with either embedding strategy for each data set and
report here the median, mean and standard deviation of the correlation
dimension estimates (computed with the values $d_e$ and $\tau$ reported
in table \ref{table}). For reference the value of correlation dimension estimated
from the time series data is also provided. All results are rounded to 3
significant figures.}
\label{local}
\end{table}

For each of these systems we estimated the best non-uniform embedding strategy
using the Genetic Algorithm and (where necessary) the sub-sample
selection scheme $30$ times. All the data sets except the longest
(the ECG recording and the laser system) produced identical results on
repeated execution. For the two longest data sets, the most often
observed embedding strategy was also the best (indicating that the sub-sample
selection scheme is expedient but perhaps not always accurate). Table
\ref{table} also illustrates that, in most cases the non-uniform embedding
covered a smaller range of embedding lags than the standard method
(i.e. $\ell_k<d_e\tau$) and is often of lower dimension
($k<d_e$). Perhaps intuitively, noisier time series required larger
$k$. Furthermore, we note that in none of the cases was the best non-uniform
embedding strategy actually uniform.

\begin{table*}
\begin{center}
\begin{tabular}{r|c|c|c|c|c|c|c}
 &\multicolumn{2}{c|} {model size} &\multicolumn{2}{c|} {prediction error} &
 \multicolumn{3}{c} {correlation dimension}\\\cline{2-8}
data & uniform & non-uniform & uniform & non-uniform & uniform & non-uniform &
data\\\hline
Sunspots & 2.07$\pm$0.828 & 2.83$\pm$0.834 & 0.472$\pm$0.0637 & 0.44$\pm$0.0523 & 1.21$\pm$0.849 & 1.15$\pm$0.614 & 1.89\\
VF & 6$\pm$1.26 & 8.6$\pm$1.79 & 0.264$\pm$0.00335 & 0.254$\pm$0.00401 & 1.01$\pm$0.492 & 1.02$\pm$0.583 & 1.62\\
Laser & 19.7$\pm$3.59 & 20.3$\pm$4.21 & 0.179$\pm$0.0166 & 0.194$\pm$0.0223 & 1.06$\pm$0.771 & 1.44$\pm$0.727 & 2.13\\
R\"ossler & 15.8$\pm$2.73 & 13.9$\pm$2.26 & 0.0353$\pm$0.00709 & 0.0395$\pm$0.00686 & 0.999$\pm$0.548 & 1.29$\pm$0.608 & 1.59\\
R.+noise & 7.53$\pm$1.48 & 6.4$\pm$0.675 & 0.0996$\pm$0.00655 & 0.103$\pm$0.00698 & 1.16$\pm$0.506 & 1.11$\pm$0.377 & 1.82\\
Lorenz & 6.77$\pm$1.04 & 12.7$\pm$3.22 & 0.131$\pm$0.0189 & 0.0597$\pm$0.00918 & 0.175$\pm$0.215 & 0.989$\pm$0.281 & 1.97\\
L.+noise & 6.3$\pm$0.988 & 6.67$\pm$1.09 & 0.159$\pm$0.00814 & 0.109$\pm$0.00861 & 0.122$\pm$0.286 & 1.03$\pm$0.311 & 1.61\\
\end{tabular}
\end{center}
\caption{Comparison of modeling results for the uniform ($d_e$ and
$\tau$) and non-uniform ($\ell_1,\ldots\ell_k$) embedding parameters listed
in Table \ref{table}. For each embedding strategy we constructed $30$
nonlinear models, with minimum description length as a selection
criterion and computed the average number of model parameters and the
average out-of-sample iterated model prediction error. For each model we
also computed the mean correlation dimension estimate (computed with the
values $d_e$ and $\tau$ listed earlier) for $30$ simulations (different
initial conditions) and report the median value over all models. For
reference the value of correlation dimension estimated from the time
series data is also provided. All results are rounded to 3 significant
figures.}
\label{result}
\end{table*}

To test how good these non-uniform embedding strategies are at modeling
the underlying dynamics, we apply two distinct modeling schemes. For
each scheme we compare the results obtained with both the uniform and
non-uniform embedding strategy. For either modeling scheme we simulate
trajectories on the underlying deterministic dynamical system in the
presence of noise. These {\em random trajectories} are iterates of the
deterministic model with the addition of random perturbations (with
expected variance less than the model mean square error) added to the
prediction at each step. In other words, if 
\[F(x_{t-\ell_1},x_{t-\ell_2},x_{t-\ell_3},\ldots,x_{t-\ell_k})=x_{t+1}+e_{t+1}\]
where $F$ is a deterministic map model with model prediction error $e_{t+1}$,
then the random trajectory $y_t$ is obtained from 
\[y_{t+1}=F(y_{t-\ell_1},y_{t-\ell_2},y_{t-\ell_3},,\ldots),y_{t-\ell_k})=y_{t+1}+\epsilon_{t+1}\]
where $y_0=x_j$ (for some $j$ selected at random) and $\epsilon_t\sim N(0,\sigma^2)$ ($\sigma_2 < E(e_t^2)$).

The first modeling scheme is essentially
iterated ``drop-one-out'' constant interpolation as described in
\cite{pps2}. By construction, the non-uniform embedding strategy will have
the optimal short term prediction. However, in Table \ref{local} we test
how well this strategy captures the long term dynamics. For each time
series and each embedding strategy we compute $30$ random trajectories
and compare the correlation dimension estimates
\cite{effgka}. Correlation dimension estimates are used here as a
quantitative comparison, we do not claim that it is an accurate (or even
unbiased) estimate of the attractor's true correlation dimension. Table
\ref{local} shows that both embedding strategies perform fairly well,
with the non-uniform embedding strategy performing significantly better for
the short or noisy data sets. In most other cases the difference is not
significant. In no instances did the non-uniform embedding strategy
perform significantly worse or fail to
capture the dynamics (the large variance for the uniform embedding
strategy indicates that it often failed to accurately capture the
dynamics of the sunspots time series).

The second modeling scheme is more sophisticated and is an attempt to
genuinely estimate the underlying deterministic dynamics of the system
(this is not possible from a local constant method, despite the
admirable results of Table \ref{local}). For each data set we compare the
uniform embedding strategy (\ref{uni}) to the non-uniform embedding
strategy (\ref{nonuni}) by constructing nonlinear models using the method
described in \cite{kJ95a}. These models are radial basis models with the
number of radial basis functions determined according to the minimum
description length principle. 

For each data set we computed $30$ nonlinear
models with either embedding strategy and report in Table \ref{result}
the average model size (the number of basis
functions in the best radial basis model) and normalized out-of-sample iterated
mean model prediction error
for the minimum description length best model (repeated
modeling attempts are required because this highly nonlinear fitting
procedure is stochastic). Furthermore, for each of the $30$ models we
generate $30$ random trajectories of $N$ time steps. For each of these iterated
predictions we computed correlation dimension using the technique
described in \cite{effgka}. Table \ref{result} also compares the
correlation dimension of the data to the simulations with either
modeling scheme. In general we observe that the non-uniform embedding scheme affords
larger models with smaller prediction errors and correlation dimensions
closer to that of the true data. That is, both the qualitative and
quantitative dynamics are reproduced much better with these non-uniform
embedding strategies. 

Time delay embedding is a fundamental technique for the reconstruction
of nonlinear dynamical systems from time series. It is commonly applied
to time series data and almost ubiquitously via estimation of $d_e$ and
$\tau$ and applying the transformation (\ref{uni}). We have argued
(based on the work of other authors) that this approach is not optimal,
and that in general one should apply a non-uniform embedding such as
(\ref{nonuni}). Currently there is no generic method for choosing the
best embedding strategy from among all possible non-uniform embeddings. The
main problem is that one must have a quantitative and easily computable
measure of the comparative suitability of competing embedding strategies
(\ref{nonuni}). Motivated by information theoretic concerns, we propose
a simple estimate of the ``goodness'' of embedding strategies based
primarily on the nonlinear prediction error of a local constant model
(\ref{dl2}). We find that it is necessary to augment this with a combination of a genetic algorithm and
sub-sample selection scheme.

After considering a wide variety of experimental and simulated time
series we conclude that this method provides alternative embedding
strategies which are often smaller ($k<d_e$ and $\ell_k<d_e\tau$) and
perform at least as well, but in general significantly better than,
standard techniques. We have applied correlation dimension as a
quantitative measure of the accuracy of dynamic reconstruction and find
that the non-uniform embedding strategy described here produces models which
behave more like the true data. 

This embedding lag selection scheme provides a method to choose good
embedding strategies for time delay embeddings. A straightforward
extension of this idea will also allow one to select variables for more
general multivariate problems. Obvious examples are in the selection
of optimal inputs for artificial neural networks and for testing
dependency among physical variables.

%\vspace{0.5cm}
\section*{Acknowledgments}

This research was supported by a Hong Kong Polytechnic University
Research Grant (NO. B-Q709). MATLAB source code is available from the author.

\bibliographystyle{unsrt}
%\bibliography{../../latex/bibliography}

\end{document}